\documentclass[prl,twocolumn,showpacs,amsmath,amssymb,floatfix]{revtex4}

\usepackage[pdftex]{graphicx} \pdfcompresslevel=9 \usepackage{color}
\usepackage[pdftex,breaklinks=true,colorlinks=tre]{hyperref}
\usepackage{bm}
\usepackage{textcomp}
\usepackage{color} \newcommand{\ie}{{\it i.e.}\ }
\newcommand{\neel}{N\'eel\ }
\newcommand{\kag}{kagom\'e\ }
\newcommand{\Zdeux}{$\mathbb{Z}_2$\ }
\newcommand\jj{$J_1-J_2$\ }
\newcommand\ratio{$J_2/|J_1|$\ }
\newcommand\mc{Monte Carlo\ }
\begin{document}
\preprint{cuboc}
\title{Chirality and \Zdeux vortices in a Heisenberg spin model on the \kag
lattice}
\author{J.-C.~Domenge$^1$}
\email{domenge@physics.rutgers.edu}
\author{C.~Lhuillier$^2$}%
\email{claire.lhuillier@upmc.fr}
\author{L.~Messio$^2$}
\author{L.~Pierre$^2$}
\author{P.~Viot$^2$}
%
%
\affiliation{$^1$Department of Physics and Astronomy and Center for
Condensed Matter Theory, Rutgers University, Piscataway, NJ
08854-8019} \affiliation{$^2$Laboratoire de Physique Th\'eorique de
la Mati\`ere Condens\'ee,\ Univ. P. \& M. Curie, \ CNRS, UMR 7600,\\
Case Courrier 121,\ 4 place Jussieu,\ 75252 Paris Cedex,\ France }

\date{\today}
%
\begin{abstract}
The phase  diagram of the classical \jj  model on  the \kag lattice is
investigated  using   extensive  \mc simulations.  In a realistic range
of parameters, this model has a low-temperature chiral-ordered  phase
without long-range spin order. We show that the critical transition
marking the destruction of chiral order is preempted by the first order
proliferation of \Zdeux point defects. The core energy of these vortices
appears to vanish when approaching the $T=0$ phase boundary, where both
\Zdeux defects and gapless magnons contribute to disordering the system
at very low temperature. This situation might be typical of a large
class of frustrated magnets. Possible relevance for real materials is
also discussed.
\end{abstract}
\maketitle
In classical spin systems, competing interactions commonly frustrate the
conventional $(\pi,\pi)$ \neel order, possibly leading to more exotic
arrangements of the local spins. Prominent examples include helicoidal
configurations,~\cite{v77,sp02,Bergman2007} which usually break space
inversion and time reversal. Noticeably, such spin chirality is a
sufficient condition for multiferroic behavior, \ie non-zero coupling
between magnetization and electric polarization\cite{Cheong2007,Ramesh2007}.

However it is not uncommon that non-planar magnetic orders relieve the
frustration even more effectively than helicoidal configurations do. The
associated magnetic order parameter is then three-dimensional, hence
also chiral. To date, two such orders have been exhibited, both on
triangular-based lattices with competing interactions: a 4-sublattice
tetrahedral order was found on the triangular lattice~\cite{mkn97} while
a 12-sublattice cuboctahedral order was found more recently on the \kag
lattice.~\cite{Domenge2005}

In two dimensions, complex magnetic orders might seem of purely theoretical
interest since gapless spin-waves destroy the spin long-range order at
arbitrarily low temperatures. However, this disordering process is soft,
in the sense that at low but finite temperature, the spin-spin
correlations remain large enough to sustain emergent long range orders.
This is examplified by the two above-mentioned models where chiral
long-range order persists up to finite temperatures, whereas long-range
order in the spins is lost. Interestingly, the emergent chiral order
parameter is Ising-like, and in a straightforward extrapolation one
expects that these chiral phases will disappear through a critical
transition in the 2D Ising universality class. However, Momoi and Kubo
showed that in the case of the tetrahedral order, this is true only in
the ``weak universality'' sense.~\cite{mkn97} 

We point  out  that such three-dimensional  magnetic orders completely
break the  $SO(3)$  symmetry of  Heisenberg interactions.   Hence  the
order parameter space is $SO(3)$ which  supports point defects, namely
vortices  in two dimensions,  around   which  the order  parameter  is
rotated by $2\pi$. However,  note that rotation of  $4\pi$ is equivalent
to the   identity, so  that  the  order parameter   may  only wind  by
$\pm2\pi$,  as        can    be  more    formally       deduced   from
$\Pi_1=\mathbb{Z}_2$. This evidences the  peculiar topology of $SO(3)$
vortices  compared to the well known  $SO(2)$ vortices involved in the
Berezinski.-Kosterlitz-Thouless (BKT)  transition In particular, since
$SO(3)$  rotations of 4$\pi$  are equivalent  to the identity, $SO(3)$
only supports vortices with unit ``winding number'', as can be deduced
more formally   through  $\Pi_1(SO(3))=\mathbb{Z}_2$.   These   \Zdeux
vortices were first exhibited in  an early numerical work by  Kawamura
and Miyashita~\cite{km85} on the antiferromagnetic Heisenberg model on
the  triangular lattice, where the  defects  were shown to proliferate
rather  abruptly at finite  temperature. To  date,  however, there  is
still  no conclusive evidence that  a  genuine phase transition indeed
takes  place in this model.  On the experimental   front, the proof of
existence of the
\Zdeux vortices remains rather elusive, although they may have been
probed indirectly in recent NMR experiments on
NaCr$_2$O$_3$.~\cite{omb06}

In this letter we exhibit \Zdeux vortices in the \jj model on the \kag
lattice and show that they are responsible  for the first order nature
of  the chiral transition and  we study the  effects of frustration on
the core energy of these defects.

The Hamiltonian of this model reads:
\begin{equation}
\mathcal{H}= J_1\sum_{\langle i,j\rangle} \vec S_i\cdot\vec S_j +
J_2\sum_{\langle\langle i,k\rangle\rangle} \vec S_i\cdot\vec S_k
\label{eq:H}
\end{equation}
where the first sum runs over pairs of nearest neighbors (at
distance 1 on the \kag lattice) and the second sum over pairs of
second nearest neighbors (at distance $\sqrt3$).

\begin{figure}[t!]
\resizebox{.305\textwidth}{!}{\includegraphics{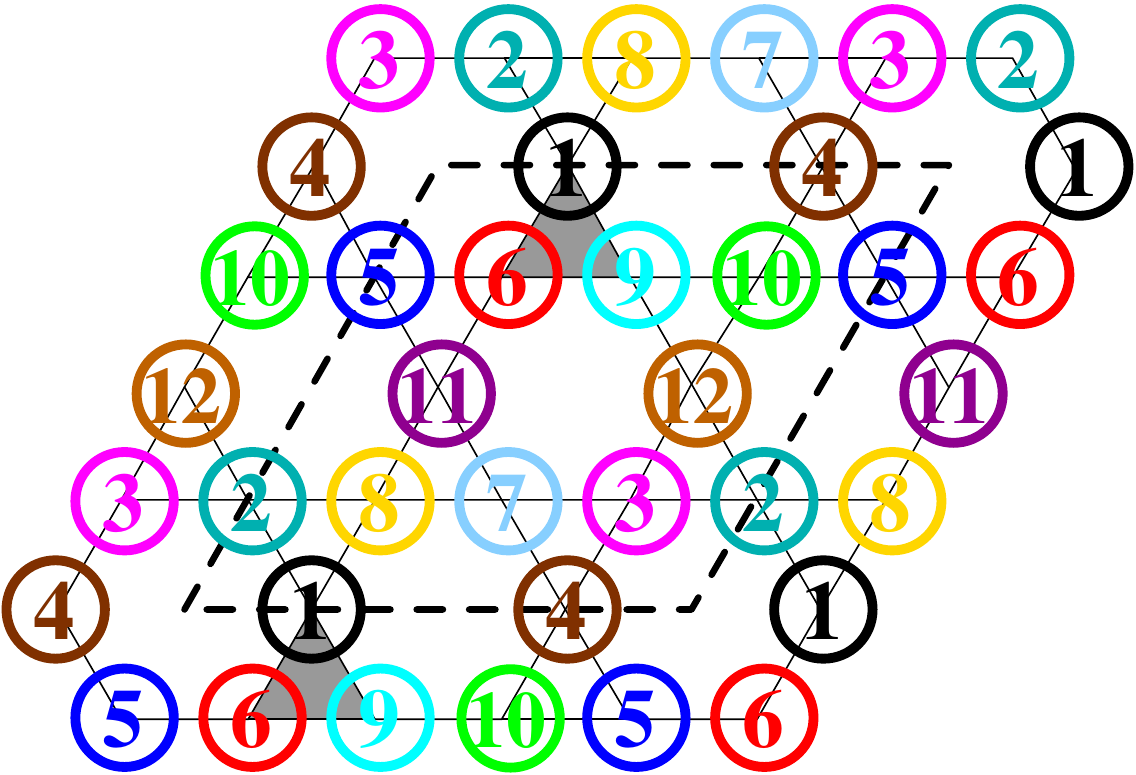}}
\resizebox{.325\textwidth}{!}{\includegraphics{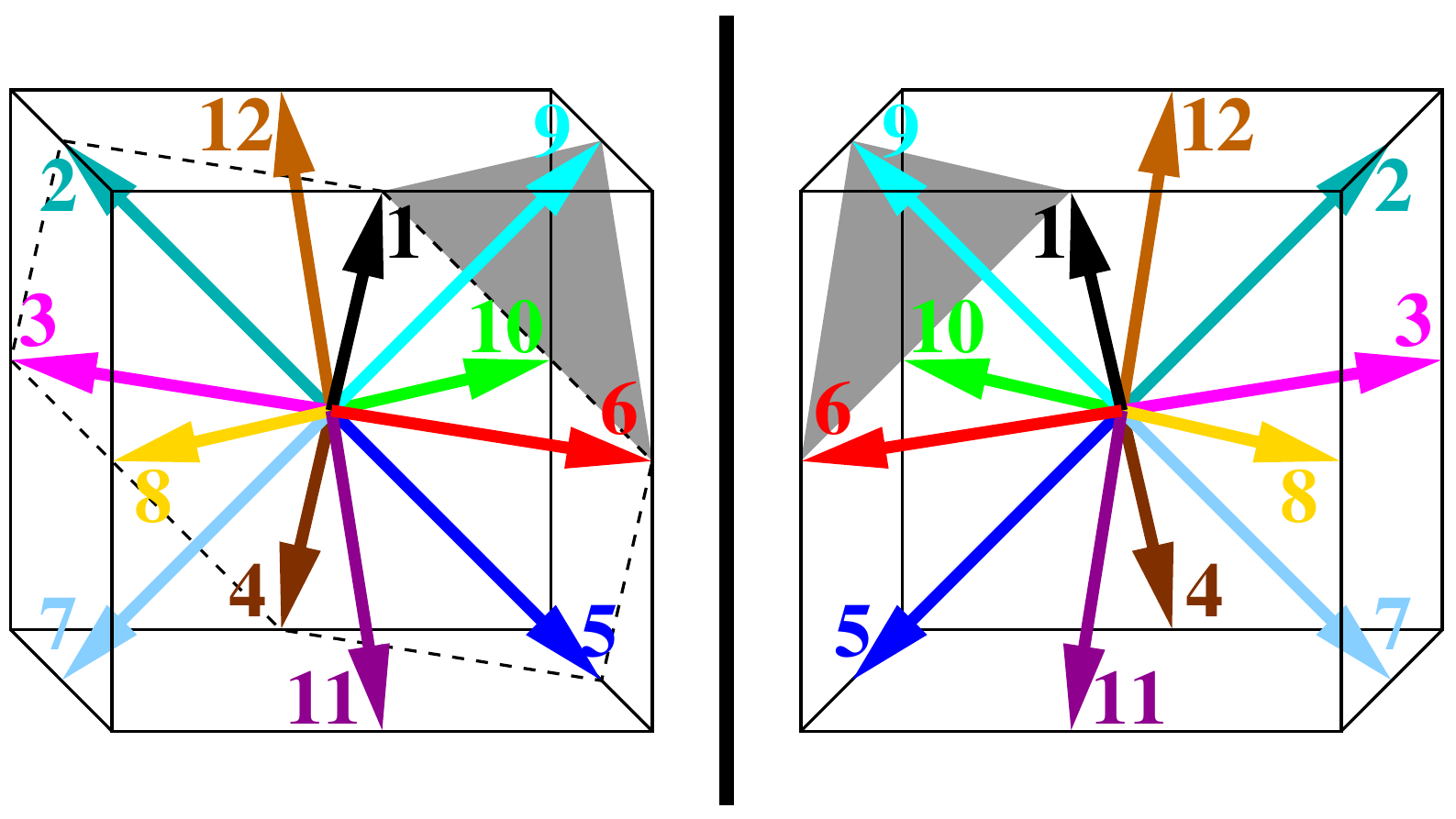}}
\caption{(Color online) 12-sublattice \neel order on the \kag lattice.
Top: The 12-site magnetic unit cell: different sublattices are indicated
by different numbers. Bottom: the order parameter in spin
space and its mirror image.  The two degenerate ground states differ
only by their scalar chirality, namely the triple product of 3 spins on
a (shaded) triangle.}
\label{fig:1}
\end{figure}

We are interested in the 12-sublattice antiferromagnetic ground state
obtained  for \mbox{$J_1<0$}  and \mbox{$J_2>|J_1|/3$}.~\cite{Domenge2005}. The
associated order parameter has the symmetry  of a
cuboctahedron, with scalar chirality $\sigma_{ijk}=\sqrt 2\;\vec
S_i\cdot\vec   S_j\times\vec S_k$ either $+1$ or $-1$, where $(i,j,k)$
label the 3 sites of a triangle clockwise (Fig.~\ref{fig:1}).
On the \kag lattice, triangles pointing up and those pointing down carry
opposite chirality in the ground state and the associated chiral long-range
order is evidenced by a finite value of the staggered chirality
\begin{equation}
{\mathcal C} =\frac{3}{2N}\sum_{\langle
ijk\rangle}(-1)^{\alpha_{ijk}}\sigma_{ijk}
\end{equation}
where the sum runs over all $2N/3$ triangles of the \kag lattice with
$\alpha_{ijk}=0$ (1) for triangles pointing up (down).

To investigate the finite-temperature  phase diagram of this model, we
perform \mc simulations using a parallel tempering algorithm.  This
method  is indeed suitable    to overcome  the   free energy  barriers
encountered at  first-order phase transitions, as we  will show is the
case for the chiral transition in this  model. Further, this algorithm
is easy to  parallelize  and yields thermodynamic quantities over a
large  range of temperatures in a single run, once combined with
reweighting methods. We simulated samples of linear size  $L$ ranging
from 12 to 64, with up  to  $N=3L^2\leq12288$ spins.  Although the
tempering method suppresses the slowing-down associated  to the crossing
of free energy   barriers, the existence of a  large  spin-spin
correlation length  drives an  ``effective critical''  slowing-down.
Therefore, for the largest samples the number of \mc steps needs to be
increased up to $2^{22}$ steps per spin.
\begin{figure}[t!]
\resizebox{7.5cm}{!}{\includegraphics{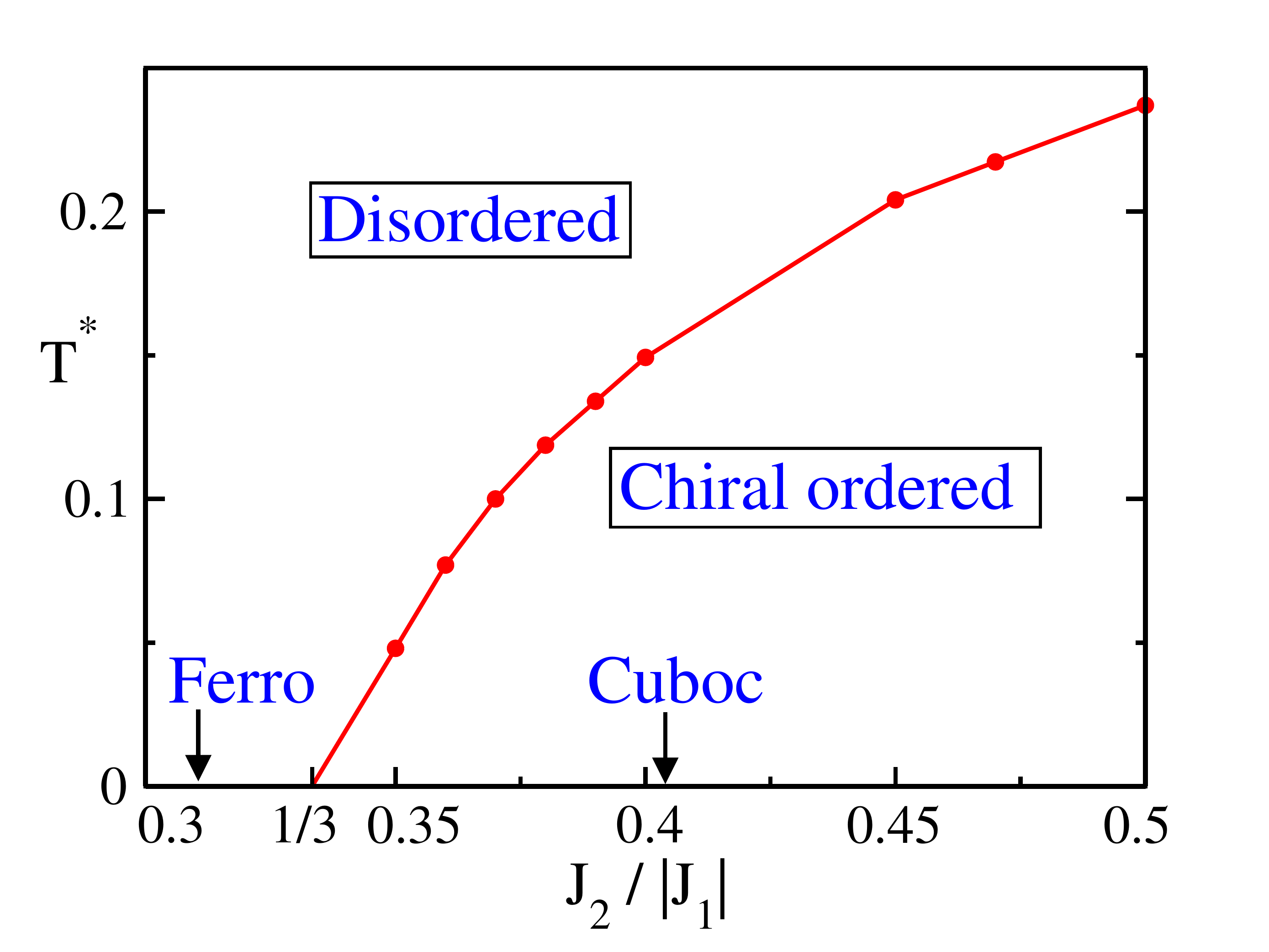}}
\caption{(Color online) Transition temperature $T^*$ from chiral order
to full disorder versus the  coupling     ratio  \ratio. For  $0\leq
J_2/|J_1|\leq1/3$, the ground-state is ferromagnetic. For  $
J_2/|J_1|>1/3$, the ground state is  the 12-sublattice \neel  ordered
phase described  in the text and abbreviated here as "cuboc". Dots:
results  of \mc simulations  (size effects are much smaller than the
size of the symbols). The solid line is a guide to the eye.}
\label{fig:2}
\end{figure}
The  first    results   of    the   simulations   are  collected    in
Fig.~\ref{fig:2}.  Starting    from \mbox{$J_2/|J_1|=1/3$},  the
temperature extent of the chiral phase increases with increasing
antiferromagnetic $J_2$. We have monitored several  thermodynamic
quantities, namely the averaged value of the energy per spin $\langle
e\rangle$, the specific heat,     $C_v/k_B=\frac{N}{(k_BT)^2}(\langle
e^2\rangle   -\langle e\rangle^2)$,   the staggered chirality
$\langle|\mathcal{C}|\rangle$ and              the associated
susceptibility
$k_B\chi_{\mathcal{C}}=\frac{2N}{3T}(\langle\mathcal{C}^2\rangle-\langle|\mathcal{C}|\rangle^2)$.
Figure~\ref{fig:3}    shows  the  rapid  destruction   of   the chiral
long-range order at the transition   (note the very small  temperature
scale) for $J_2/|J_1|=0.38$.

To characterize the   chiral  transition we use standard   finite-size
scaling analysis.  For  $1/3<J_2/|J_1|<0.45$,  the energy distribution
becomes bimodal in the neighborhood of the  transition, and its maxima
become   more pronounced with  increasing   sample sizes.  Both the
maximum of  the specific heat  $C_v^{\rm  max}(L)$ and  of the  chiral
susceptibility  $\chi_\mathcal{C}^{\rm max}(L)$ increase algebraically
with     $L$      with    exponents     within    $2.0\pm0.15$     for
$J_2/|J_1|\leq0.38$. For $J_2/|J_1|=0.39$ and 0.40, the scaling regime
is  reached   only for  the largest  samples.    In addition, denoting
$T_{C_v}(L)$ and   $T_{\chi_\mathcal{C}}(L)$ the temperatures   of the
maximum   of     $C_v(L)$  and   $\chi_\mathcal{C}(L)$    one  obtains
$1/T_{C_v,\chi_\mathcal{C}}(\infty)-1/T_{C_v,\chi_\mathcal{C}}(L)\sim1/L^2$
for $J_2/|J_1|\leq0.38$. The  above analysis shows that the transition
does not belong   to  the expected Ising  universality  class,  but is
first-order, although the  increasing difficulty to  reach the scaling
regime, when increasing \ratio,  evidences the concomitant growth of a
correlation  length.  This is consistent with  the continuous decrease
of the latent  heat   with increasing \ratio ($0.034, 0.028, 0.012$ for
$J_2/|J_1|=0.36,0.38,0.45$)

To account  for the discontinuous  nature  of the chiral  transition a
thermally activated mechanism was looked for: proliferation of point
defects is an obvious candidate, and we now proceed to compute the
\Zdeux vortices of the model. We define the local trihedron of a
12-sites magnetic unit cell as follows: $\vec e_1=(\vec  S_i+\vec
S_j)/|\vec S_i+\vec S_j|$, $\vec e_3=(\vec S_i-\vec S_j)/|\vec S_i-\vec
S_j|$ and $\vec e_2=\vec e_3\times\vec e_1$, with $i$ and $j$  any two
sites in the magnetic unit cell  that carry non-collinear spins. These
trihedra live on  a triangular super-lattice of   spacing  $4$ in  units
of the original \kag lattice. Then we determine the rotation of axis
$\vec n$ and angle $\theta$ between  two successive trihedra, and
disambiguate it from $(-\vec n,2\pi-\theta)$ by  using  its universal
covering element $U(\vec
n,\theta)=e^{-i\frac{\theta}{2}\vec\sigma\cdot\vec n}\in SU(2)$, where
$\vec\sigma$  are  the three  Pauli  matrices. Finally we define the
\Zdeux vorticity inside an elementary (triangular) contour  by
\begin{equation}
V_\bigtriangleup=  \frac{1}{2}\left(1-\frac{1}{2} Tr\left(\prod_{j\in
\bigtriangleup}U(\vec n_j,\theta_j)\right)\right)
\end{equation}
where $V_\bigtriangleup=1(0)$ when the loop $\bigtriangleup$ encloses a
singularity (otherwise). 
\begin{figure}[t]
\resizebox{8cm}{!}{\includegraphics{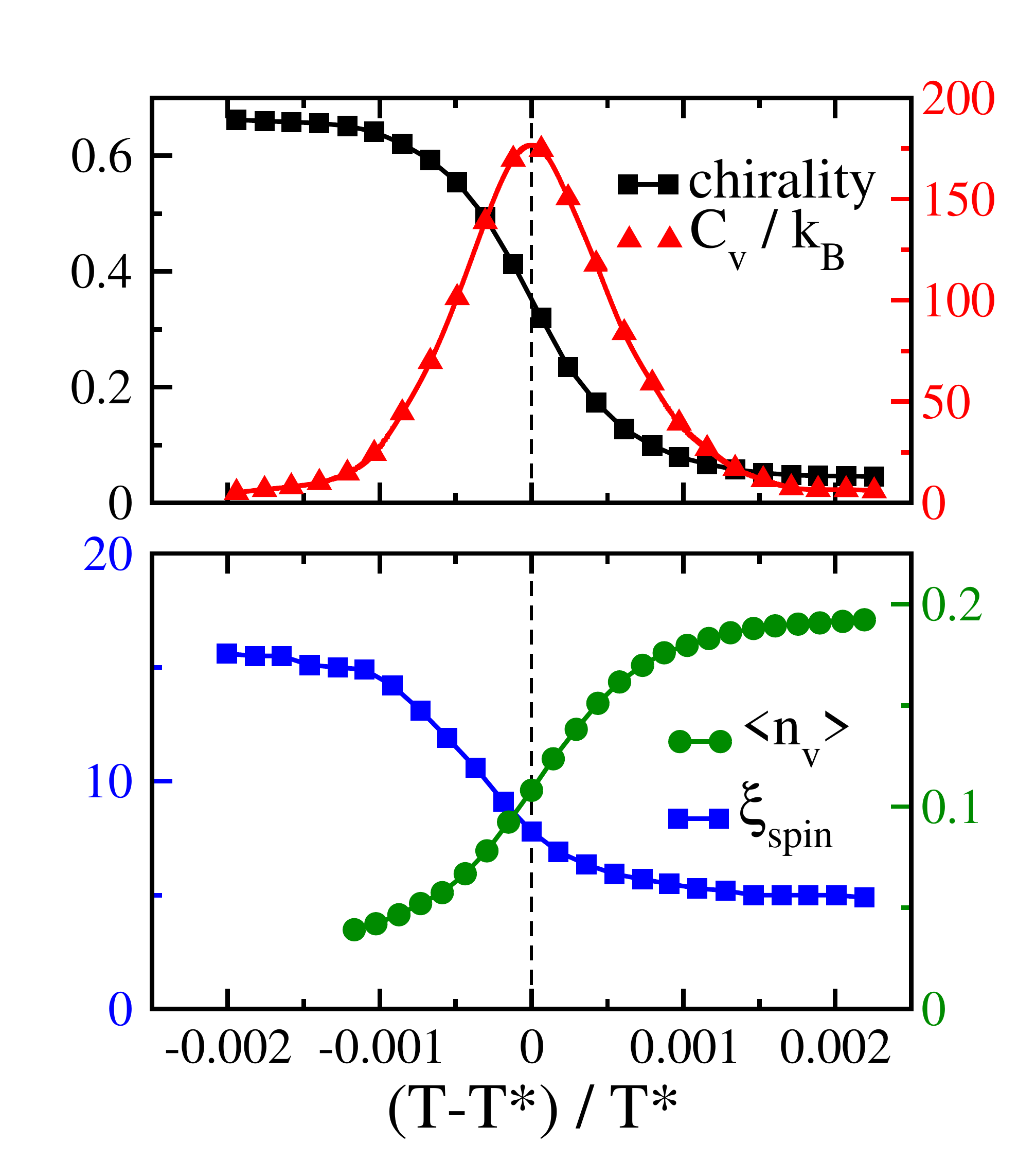}}
\caption{(Color online) Top: Temperature dependence of the staggered
chirality $\langle|\mathcal{C}|\rangle$ and the specific heat for
$J_2/|J_1|=0.38$ and $L=64$. Temperatures are measured relatively to
$T^*$, temperature where $C_v$ is  maximum. Bottom: spatially averaged
spin-spin correlation length $\xi_{\rm spin}$  and the vortex density
versus temperature.}
\label{fig:3}
\end{figure} 
We      computed           the      vortex     density $\langle
n_v\rangle=6/N\sum_\bigtriangleup  V_\bigtriangleup$,  where the  sum
runs  over all $N/6$  triangles of the  super-lattice. The results are
displayed in Fig.~\ref{fig:3}  versus $(T-T^*)/T^*$, where $T^*(L)$ is
the temperature of the maximum of $C_v(L)$. They clearly show that the
chiral  transition      is concomitant   with   the proliferation of
vortices.

The behavior  of the spin-spin correlation length  $\xi_{\rm spin}$
close to  the  transition evidences that vortex proliferation
drastically decreases the  magnetic short  range order  as well as it
kills  the emergent chiral order, far before the expected critical
regime of the chiral phase is attained. Hence the Ising  chiral
transition  is avoided simply because the vortex  proliferation triggers
a first-order phase transition that preempts the critical regime. We
emphasize that the disordering effect of the \Zdeux defects is much
stronger than that of $\mathbb{Z}$ vortices at the BKT
transition.~\cite{IWMSK06} Fig. 3 indeed shows that $\xi_{\rm spin}$ is
divided by three in a temperature range $\sim 10^{-3}\,T^*$, and this
brutal decrease may even be smoothed by the finite size of the sample.
 
\begin{figure}[t]
\resizebox{8cm}{!}{\includegraphics{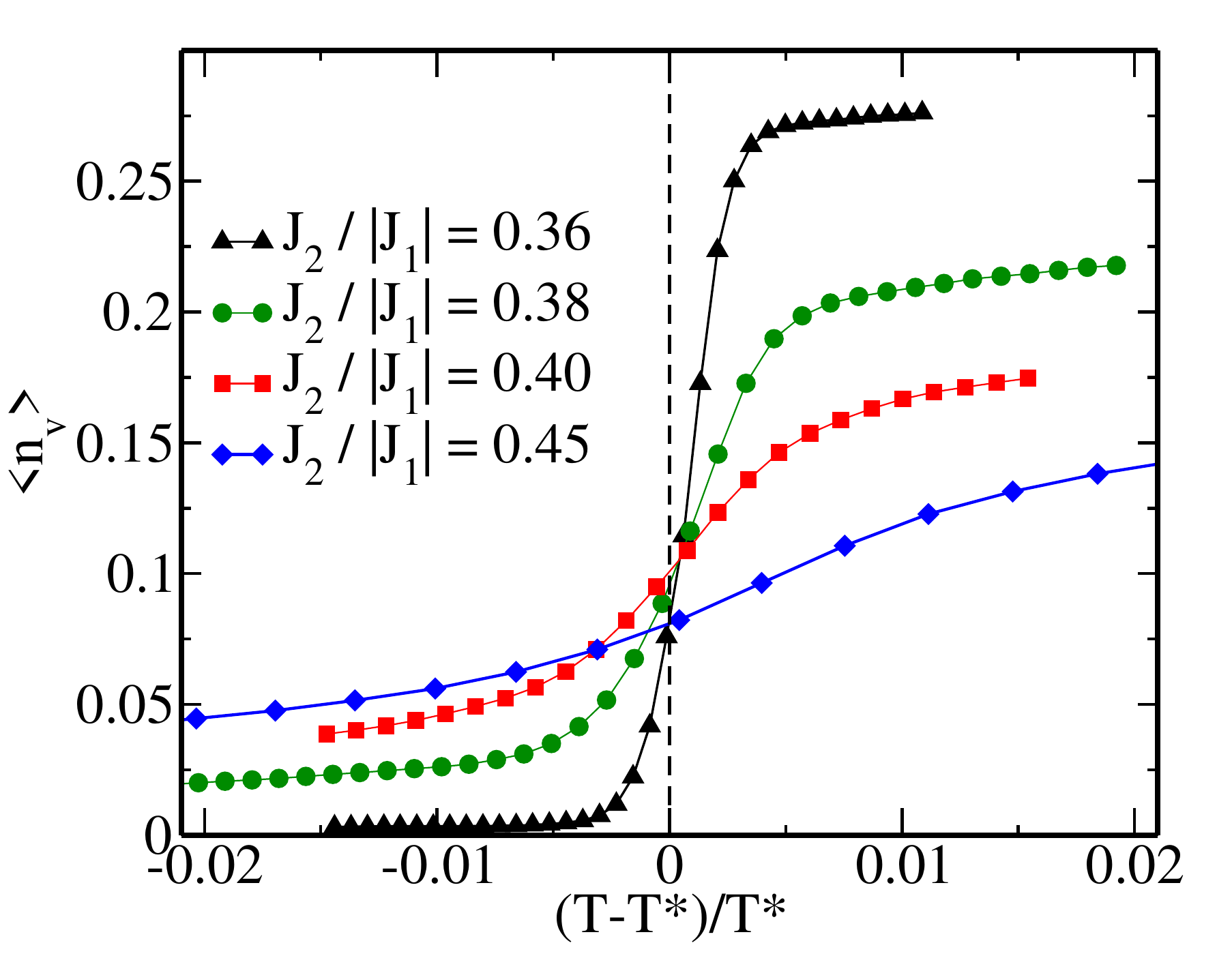}}
\caption{(Color online) Evolution of the vortex density $\langle
n_v\rangle$ around the transition temperature $T^*$ as a function of
\mbox{$J_2/|J_1|$.}}
%
\label{fig:4}
\end{figure}

The chiral correlation length $\xi_{\rm chiral}$, as computed from the
structure factor of the chirality $\mathcal{C}$ at the transition,
measures the discontinuity of the chiral transition. For $J_2/|J_1|=
0.38$ we find $\xi_{\rm chiral}=6$, while it exceeds the largest
available lattice size ($\xi_{\rm chiral}>64$) already for
$J_2/|J_1|=0.40$.  Consistently,  the vortex  proliferation smoothes
upon increasing \ratio (Fig.~\ref{fig:4})   and for  large enough
$J_2/|J_1|=0.45$, the number of   vortices at the transition is  clearly
seen  to decrease.  Although this is not the core of our study, note
that the fast growth of $\xi_{\rm  chiral}$ with increasing \ratio makes
it particularly   delicate  to   idenitfy a possible critical end point
to the line of first order chiral transitions shown in
Figure~\ref{fig:2}. 

{\it Discussion:} The chiral transition in the present \jj model may be
typical  of  many  frustrated magnets.  Indeed, as emphasized above, as
long as  the ground  state completely breaks the $O(3)$ symmetry of the
Heisenberg Hamiltonian, as is the case for non-planar \neel orders, both
chirality and \Zdeux vortices exist. The complete breaking of $SO(3)$
induces \Zdeux point defects, while the breaking of time reversal (the
discrete part of $O(3)$) leads to chiral degenerate ground-states. Hence
we expect very similar features for the chiral transition in the present
model and that observed on the triangular lattice by Momoi {\it et
al.}:~\cite{mkn97} in a posterior study, these authors indeed noticed that
the chiral transition may be weakly discontinuous.

However, the nature of the chiral transition cannot be deduced from
symmetry arguments alone and will ultimately depend on the energetics of
the two competing mechanisms that suppress chiral order: formation of
chiral domain walls versus creation of \Zdeux vortices.

In this respect, the case of the $J_1-J_3$ model on the square lattice
is interesting.  Capriotti and Sachdev~\cite{cs04b} have shown that in
this model, the doubly degenerate helicoidal ground-state results in a
finite temperature chiral phase. Although \Zdeux vortices are allowed
by symmetry, the chiral transition at finite temperature clearly falls
in the 2D-Ising universality class.~\cite{cs04b} Consistently, we
understand that in a problem dominated by antiferromagnetic
long-wavelength fluctuations, forming a chiral domain wall is much less
costly than creating a \Zdeux vortex.

Correspondingly,  in the  \jj  model   under study,   we  expect  that
competing ferromagnetic   and    antiferromagnetic interactions   will
enhance  the  short-range   fluctuations, hereby decreasing   the core
energy of \Zdeux vortices.  This  qualitative argument is supported by
the  observation that the  discontinuity of  the chiral transition  is
maximal when the competition of interactions is  highest, \ie close to
$J_2/|J_1|=1/3$: there the chiral  transition is clearly triggered  by
the proliferation of defects.

Note also that the decrease of the transition temperature as
$J_2/|J_1|\to1/3$ suggests that the core energy of the \Zdeux vortices
vanishes at the $T=0$ phase boundary, \ie  $J_2/|J_1|=1/3$.  

However, in this very low temperature regime quantum fluctuations are
expected to play a significant role. The $T=0$ quantum problem was
actually studied previously by three of us:~\cite{Domenge2005} using
exact diagonalizations for spins 1/2 as well as spin-wave expansions, it
was shown that both classical ground states, ferromagnetic and
12-sublattice antiferromagnetic, survive quantum fluctuations with a
quantum phase transition located at $J_2/|J_1|\simeq1/3$. Hence,
although the present study is purely classical, it yields the promising
prospect that additional excitations become gapless exactly at the
quantum phase transition, supplementing the usual gapless magnons that
exist in either of two neighboring phases. This is highly reminiscent of
the breakdown of the Landau paradigm in quantum spin
systems,~\cite{Senthil2004} although the route from classical \Zdeux
vortices to some fractionalized ``spinon'' excitations for spins 1/2
remains a totally open problem.

Experimentally, a large number of magnets on the \kag lattice have been
synthetized up to now, and it has been a long route to arrive at the
Herbertsmithite Z$_n$CuO, which remains an antiferromagnetic spin liquid
down to 50\,mK,\cite{Helton2007,bert:117203} with a dominant
antiferromagnetic interaction of about 190\,K. Amongst other
difficulties, the sign of the nearest neighbor coupling is a pending
problem: Atacamite, parent both of Paratacamite and Herbertsmithite,
becomes ferromagnetic around 10\,K.
%
%
Similarly, Cutitmb, the organic compound at the origin of our interest
in the present model,~\cite{hky02,nkhdslskk04} was recently shown to
experience a transition to three dimensional ferromagnetic order around
500\,mK~\cite{katsumata}: this tendency to ferromagnetism is deeply
rooted in the geometry of the exchange paths between nearest neighbor Cu
ions on the \kag lattice. If the malediction of such low-temperature
ferromagnetic orderings can somehow be avoided, possibly through a
decrease of the inter-layer couplings, the present study would be of 
direct experimental interest, beyond its initial theoretical motivation.

In  this letter, we showed that in the \jj model on the \kag lattice,
chiral order is wiped out at finite temperature by the first-order
proliferation of \Zdeux vortices. In the region of extreme frustration
$J_2/|J_1|\simeq1/3$ the core energy of the defects decreases and
appears to vanish exactly at the $T=0$ phase boundary.  This behavior is
probably common to frustrated spin systems in which competing
interactions lead to three-dimensional antiferromagnetic order
parameters.

We acknowledge important discussions with D. Mouhanna and B.  Delamotte
on  the  issue of  universality     in frustrated magnets,  and   with
M. Mostovoy and D. Khomskii\cite{bbmd07} on orbital order. C.L and L.M.
acknowledge hospitality in KITP.  This  research was supported in part
by the National Science Foundation under Grant No. PHY05-51164.


\end{document}